\documentstyle[twoside,fleqn,espcrc2,epsf]{article}

\newcommand{\ewxy}[2]{\setlength{\epsfxsize}{#2}\epsfbox[40 40 530 530]{#1}}

\newcommand{\beq}{\begin{equation}}
\newcommand{\eeq}{\end{equation}}
\newcommand{\beqa}{\begin{eqnarray}}
\newcommand{\eeqa}{\end{eqnarray}}
\newcommand{\pslash}{\not\!p}
\newcommand{\qslash}{\not\!q}

\newcommand{\Tr}{{\rm Tr}\,}

\title{The Running Coupling from the Quark--Gluon Vertex}

\author{Jon Ivar Skullerud
\address{CSSM and Department of Physics and Mathematical Physics, The
University of Adelaide, Adelaide, SA 5005, Australia {\em and} UKQCD
Collaboration}}

\begin{document}
\pagestyle{empty}

\begin{abstract} 
We present results for the QCD running coupling obtained from
measuring the quark-gluon vertex in Landau gauge with suitable
renormalisation conditions. The issue of discretisation errors arising
from the fermion action is discussed.
\end{abstract}

\maketitle


\section{Introduction}

The quark--gluon vertex is one of the fundamental quantities of QCD,
and studying the form of the vertex can give us greater insights into
the dynamics of the theory and may provide and important necessary
input into Dyson--Schwinger equations. Here we will focus on using the
quark--gluon vertex to extract the running coupling from first
principles.

\section{The vertex function}

The full (unamputated) momentum space vertex function can be defined as
\beq
G^a_\mu(p,q)^{ij}_{\alpha\beta}=\left<S^{ij}_{\alpha\beta}(p)A^a_\mu(q)\right>
\eeq
and the amputated (OPI) vertex function
\beqa
\lefteqn{\Lambda_\mu^{a,\rm lat}(p,q) =} \nonumber\\
& &\!\!\left<S(p+q)\right>^{\!-\!1}\!
\left<S(p)A^a_\mu(q)\right>\!\left<S(p)\right>^{\!-1}\left<D(q)\right>^{\!-\!1}
\eeqa
$D(q)$ is the scalar part of the gluon propagator, given in the Landau
gauge by
\beq
D_{\mu\nu}^{ab}(q) = \delta^{ab}T_{\mu\nu}(q)D(q)
\eeq
where $T_{\mu\nu}$ is the projection onto transverse fields.

Using the requirements of Lorentz covariance and parity and charge
conjugation invariance, one can derive the following general form for
the vertex in the continuum:
\begin{eqnarray}
\lefteqn{\Lambda_\mu(p^2,q^2,pq) \equiv 
\frac{1}{N_C^2-1}\Tr_{col}T^a\Lambda^a_\mu(p,q)}\nonumber \\
 & = & F_1 p_\mu + F_2 q_\mu + F_3 \gamma_\mu \nonumber \\
 & & + F_4 \pslash p_\mu + F_5 \pslash q_\mu + F_6 \qslash p_\mu 
+ F_7 \qslash q_\mu \label{eq:formfactors}\\
 & & + F_8 \sigma_{\mu\nu} p^\nu
+ F_9 \sigma_{\mu\nu} q^\nu \nonumber \\
 & & + F_{10}\epsilon_{\mu\nu\kappa\lambda}\gamma_5
\gamma^{\nu}p^{\kappa}q^{\lambda} \nonumber
\end{eqnarray}
where all the $F$'s depend only on the invariants $p^2, q^2$ and $pq$.

At tree level, this reduces to $\Lambda^0_\mu =
\frac{i}{2}g_0\gamma_\mu$.  From this we can see that the form factor
containing the running coupling is $F_3$, while all the other form
factors are expected to be finite.

If we define
\beq
K_{\mu}(p,q) \equiv i\Tr\gamma_{\mu}\Lambda_{\mu}(p,q)
\eeq
we find that $iF_3(p^2\!,\!0,\!0)=K_{\mu}(p,\!0)|_{p_\mu=0}$.

This kinematics can then be used to define a momentum subtraction
scheme for the renormalised coupling:
\beq
g_R^{\!MOM}(\mu) = -2i Z_\psi\!(\mu)Z_A^{1\!/\!2}\!(\mu)
F_3(p^2\!,\!0,\!0)|_{p^2=\mu^2}
\label{eq:gr}
\eeq
where $Z_\psi$ and $Z_A$ are the renormalisation constants for the quark
and gluon fields, defined in the Landau gauge by
\beqa D(p^2)|_{p^2=\mu^2} & = & Z_A(\mu)\frac{1}{\mu^2} \\ \Tr
(\gamma\tilde{p})S^{-1}(\tilde{p})/{\tilde{p}}^2|_{p^2=\mu^2} & = &
\frac{i}{Z_\psi(\mu)} \eeqa
$\tilde{p}_\mu=\frac{1}{a}\sin k_{\mu}a$ is used in the definition of $Z_\psi$ to
make it more `continuum-like'.

$g_R^{MOM}$, as defined in (\ref{eq:gr}) can then be related
perturbatively to the running coupling calculated in other schemes,
eg.\ $g_R^{\overline{MS}}(q^2)$.  This matching can be performed
entirely within continuum perturbation theory \cite{nprenorm}.

The running of $g_R^{MOM}$ can also be compared to the perturbative
continuum coupling derived from the two-loop beta function
\beq
g^2(\mu) = \left [ b_0 \ln(\mu^2/\Lambda^2) +
\frac{b_1}{b_0}\ln\ln(\mu^2/\Lambda^2) \right ]^{-1}
\label{eq:beta-2loop}
\eeq
with $b_0=11/16\pi^2$, $b_1=102/(16\pi^2)^2$, by doing a one-parameter
fit to this formula.
%

\section{Computation and results}

332 quenched configurations have been analysed at $\beta\!=\!6.0$
($a^{\!-1}\!=\!2$GeV) with a lattice size of $16^3\!\!\times\!\!48$.  The
propagators have been generated using the tadpole improved
Sheikholeslami--Wohlert action, at $\kappa\!=\!0.137$. The gauge fields
and propagators have been fixed to Landau gauge, with accuracy
$\frac{1}{VN_c}\sum_{x,\mu} \left |\partial_\mu A_\mu(x)\right|^2 <
10^{-12}$.

The full vertex $\gamma_{\mu}G_{\mu}(p,q)$ has a clear, symmetric
signal for $q\!=\!0$ and all values of $p$ where $p_\mu=0$
\cite{Melb}.  For $q \neq 0$ the signal falls off rapidly, and
disappears entirely for $|qa| \geq \frac{\pi}{4}$.  This makes it
difficult to implement a renormalisation scheme where $q\neq 0$,
although this would be preferable.



\begin{figure}[htb]
\ewxy{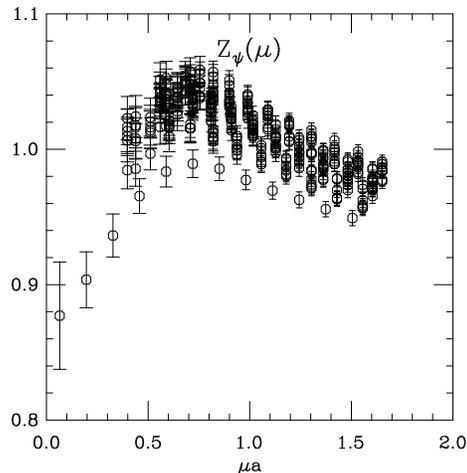}{190pt}
\vspace{-12mm}
\caption{$Z_\psi(\mu)$ as a function of $\mu$, for 80 configurations.}
\label{fig:zpsi}
\end{figure}

In order to compute the proper vertex, one needs the quark
renormalisation constant $Z_\psi$.  As figure \ref{fig:zpsi} shows,
this suffers from serious ambiguities, especially at high momenta.  A
more detailed analysis shows these ambiguities to be a result of
violation of rotational symmetry, which can be attributed to $O(a)$
errors in the fermion action: although the SW action is $O(a)$
improved, this is only the case for on-shell, gauge invariant
quantities.  In order to remove $O(a)$ errors for gauge dependent,
off-shell quantities, one requires two additional counterterms in the
actions \cite{Sharpe97}.

\subsection{The proper vertex and $g_R$}

One feature of the data is the strong correlations between data at
different momenta.  This makes an assessment of possible violations of
rotational and reflection symmetry difficult, as it turns out that
different samples of 83 configurations can have values of $K_{\mu}$ as
much as $3\sigma$ apart.  With this proviso, we find that for
$q\!=\!0$ and $p_{\mu}\!=\!0, K_{\mu}(p,q)$ is independent of $\mu$ within
errors, as one would expect.  For $p_{\mu} \neq 0$, the $F_4$ form
factor appears, and this has a large effect on $K_{\mu}$ --- making it
consistent with 0 for all but the smallest values of $p_\mu$.

\begin{figure}[htb]
\ewxy{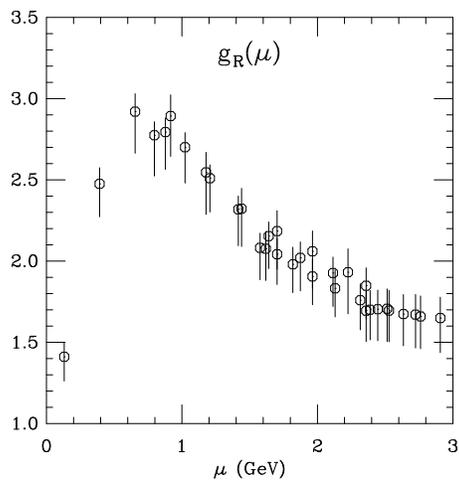}{190pt}
\vspace{-12mm}
\caption{$g_R^{MOM}(\mu)$ as a function of $\mu$.}
\label{fig:gr}
\end{figure}

The renormalised coupling is showed as a function of $\mu$ in figure
\ref{fig:gr}.  The signal is encouragingly clean, and it exhibits
qualitatively the same behaviour as the running coupling extracted
from the 3-gluon vertex \cite{ggg}.

\begin{figure}[htb]
\ewxy{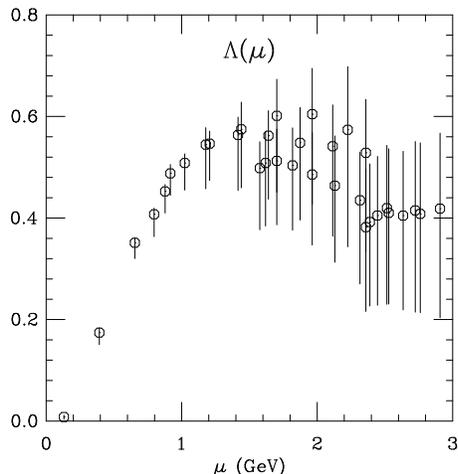}{190pt}
\vspace{-12mm}
\caption{$\Lambda^{MOM}$ as a function of the renormalisation scale
$\mu$.} 
\label{fig:lambda}
\end{figure}

An estimate for $\Lambda_{QCD}$ in this scheme has been obtained by
inverting equation (\ref{eq:beta-2loop}), and the result is plotted as
a function of momentum in figure \ref{fig:lambda}.  The running of
$g^{MOM}_R$ will be consistent with two-loop continuum perturbation
theory where this estimate is consistent with a constant.  However, no
firm conclusions can be drawn in this case, since the point where the
data becomes consistent with a constant value for $\Lambda$ (above
1GeV) is also where noise begins to dominate.

\section{Discussion and conclusions}

The feasibility of using the quark--gluon vertex to extract the QCD
running coupling has been demonstrated, although $O(a)$ errors in the
quark propagator remain a problem.  This may be addressed by using an
off-shell improved fermion action, but further work is needed to
investigate the feasibility of this.

These results will be matched perturbatively to the ${\overline{MS}}$
scheme to enable a quantitative comparison with other determinations
of $\alpha_s$.  A study of the other form factors in the vertex, as
well as the possible effects of Gribov copies, is also underway.

\section*{Acknowledgements} 

This work has been supported by Norwegian Research Council grant
100229/432, the Australian Research Council, and EPSRC grants
GR/K41663 and GR/K55745.  I wish to thank Claudio Parrinello, David
Richards and Tony Williams for fruitful discussions.

\end{document}